# LiDAR Accuracy on North American Mountain Summits

Eric Gilbertson[1], Richard Hensley[2], Andrew Kirmse[3], Kyle Bretherton[4], Kathryn Stanchak[5]

1. Department of Mechanical Engineering, Seattle University, Seattle, WA, USA

2. Las Vegas, NV, USA

3. Redwood City, CA, USA

4. Seattle, WA, USA

5. Issaquah, WA, USA

**Abstract**

Mountainous terrain is increasingly being measured and mapped by airplane-based LiDAR (Light Detection and Ranging) techniques, but the accuracy of these measurements in such topographically variable terrain is not well understood. For this study we measured 179 mountain summits with differential GNSS static surveys and compared summit elevation and location measurements to those measured by LiDAR in point cloud data sets. We measured summits in 13 US states (Washington, Idaho, Montana, Utah, California, Nevada, Arizona, New Mexico, Michigan, Wisconsin, Kentucky, Colorado, and Pennsylvania) and two Canadian provinces (British Columbia and Nova Scotia). Summits included icecapped peaks, open rocky peaks, and tree-covered peaks ranging in elevation from 490m to over 4000m. LiDAR-point-cloud-derived summit elevations and locations were computed using four different methods: manual processing, highest ground return, highest return, and Lastools reclassification. The average one-sigma LiDAR vertical errors for each method were 0.50m, 1.09m, 9.83m, and 1.96m, respectively. Average one-sigma horizontal errors were 3.03m, 2.41m, 5.17m, 3.78m, respectively. Errors are also presented separately for each type of summit. Error sources include sharp summits being unsampled, dense vegetation misclassified as ground, human-created structures misclassified as ground, snow/ice that can melt over time, and summit erosion over time.

**Introduction**

Approximately 24% of land on Earth is categorized as mountainous terrain (Korner 2021), and it is important to have accurate elevation data for mountains in these regions. For instance, accurate maps for navigation rely on accurate mountain elevations. Knowing accurate mountain elevations allows for water management planning in watershed areas (Wu 2007, Dennedy-Frank et al. 2024), and many climate models depend on accurate mountain elevations (Mountain Research 2015). Some mountains with permanent icecaps at their summits are losing elevation due to climate change (Gilbertson et al. 2025). Volcanic peaks can change elevation before eruptions (Eggers 1983). Finally, local tourism and recreation industries in mountainous areas often rely on knowing the highest peaks in a region and want to give interested visitors the most accurate elevation measurements.

Mountain summit elevations in the United States and Canada were first measured using trigonometric surveys starting in the late 1800s and early 1900s. Many of these initial measurements were updated with photogrammetry in the late 1900s, though vertical accuracies were generally not reported (Usery and Varanka 2009). Starting in the late 1980s, a few mountain peaks like Mt. Rainier (Signani 2000) began to be surveyed with differential GNSS (dGNSS), which can have vertical accuracy as low as 2cm (Shao and Sui 2015) but are time-consuming and labor-intensive to conduct. Accuracies were not reported for the Mt. Rainier dGNSS surveys.

In the early 2000s, the US and Canada began conducting airplane-based LiDAR (Light Detection and Ranging) surveys to update topographic maps and mountain elevation data with increased and known accuracy over triangulation- and photogrammetry-based surveys (Crane et al. 2002). In general, in the US and Canada only one LiDAR survey has been conducted in each mountainous region, though some significant mountains (like Mt. Rainier and Mt. St. Helens) have had multiple LiDAR surveys. Other countries in Europe (Ada 2024) and the Pacific Islands (Wandres 2024) have also conducted airplane-based LiDAR topographic surveys.

For an airplane-based LiDAR survey, the airplane's elevation and coordinates are known with high accuracy (horizontal one sigma errors 5-15cm, vertical 7.5-22.5cm, May 2007) from onboard GNSS units and inertial measurement units. The plane sends a pulse of light to the ground and measures the time of return to the plane. This time of light travel can be used to calculate the distance from the plane to the ground, which can then be used with the plane's known position to calculate the elevation and latitude and longitude coordinates of the sampled location on the ground.

Sampled points in a LiDAR survey have nominal horizontal spacing 0.35m (USGS 2025) for US and Canadian datasets. However, in practice we have found horizontal spacing can be up to 2m in the mountainous terrain studied. Locations between samples are unmeasured.

Elevation models can be used to extrapolate measurements to these unsampled spaces, but the reliability of these models depends on the smoothness of the terrain; abrupt topographical changes between the measured points can still result in errors. Thus, the accuracy of LiDAR relies on the accuracy of the measured data points.

The accuracy of LiDAR point cloud measurements has been studied in flat terrain including in forests (Gillin et al. 2015), open fields (Kucharczyk et al. 2018), salt marshes (Liu 2022), parking lots (Elaksher and Alharthy 2023) and sand dunes (Schmelz and Psuty 2019). Vertical errors ranged from 0.12m – 0.52m (Table 1). The areas with the lowest errors tended to be very flat and open parking lots or fields, while areas with higher errors generally had more dense vegetation or the terrain was less uniform.

Fewer studies have measured LiDAR accuracy in mountainous terrain, possibly due to increased challenges in accessing these remote areas with dGNSS surveying equipment. Studies have been conducted in alpine terrain in Slovakia (Kovanič et al. 2020) and on several mountain summits in Poland (Maciuk 2021), with vertical errors ranging from 0.2m – 0.75m. The higher errors may be a result of a combination more varied topography and small sample sizes. For instance, only six summits have data where static dGNSS measurement were compared to LiDAR point cloud data.

Table 1: LiDAR accuracy studies 2018 – 2023 categorized by terrain type, ground coverage, model used, and reported vertical error.

| Study | Region | Terrain | Cover | Model | Vertical Error (m) |
| --- | --- | --- | --- | --- | --- |
| Kucharczyk et al. (2018) | Alberta, Canada | Flat | Open | Point Cloud | 0.24 |
| Schmelz and Psuty (2019) | New York, USA | Flat | Trees | Point Cloud | 0.50 |
| Kovanič et al. (2020) | Slovakia | Mountainous | Open | Point Cloud | 0.75 |
| Maciuk (2021) | Poland | Mountainous | Open | Point Cloud | 0.2-0.4 |
| Liu (2022) | California, USA | Flat | Trees | Point Cloud | 0.52 |
| Elaksher and Alharthy (2023) | New Mexico, USA | Flat | Open | Point Cloud | 0.12 |

In this paper we focus on the specific topic of the accuracy of LiDAR point cloud data (i.e., non-modeled data) relative to static dGNSS measurements in determining the elevations of mountain summits. This compares the most accurate ground survey method to the most accurate LiDAR summit elevation measurements. Mountain summits pose several unique challenges for LiDAR measurements (Figure 1).

Sharp summits, where the elevation varies by at least the nominal vertical error of the LiDAR collection in the average sample spacing, can be difficult to measure. An example is Mt. Davis (Fig 2). If a summit is sharp, it is unlikely the LiDAR measurement will hit the exact summit due to the 1-2m gap between measured points, which can mis abrupt topographical changes. Thus, sharp summits will likely be under-measured by LiDAR.

Summits with dense brush or trees are prone to over-measurement by LiDAR. If a sampled point of dense brush is misclassified as ground, then this gives an erroneously high measurement for the summit elevation.

Mountain summits are sometimes covered in permanent ice or snow that persists through the summer. These summits can melt over time, as has been measured on Mt. Rainier (Washington, USA; Gilbertson et al. 2025). Thus, an outdated LiDAR measurement may be erroneously high. Similarly, some mountains can erode over time, such as Mt. St. Helens (Washington, USA; results presented in this paper). In this case, an older LiDAR measurement may also give an erroneously-high summit elevation.

Finally, artificial human-made structures on summits can be misinterpreted as natural features in LiDAR data. This is relevant for human-made cairns and concrete structures that may be classified as ground in LiDAR data.

With this study we aim to measure the accuracy of mountain summit elevations and locations that are derived from LiDAR point cloud data.

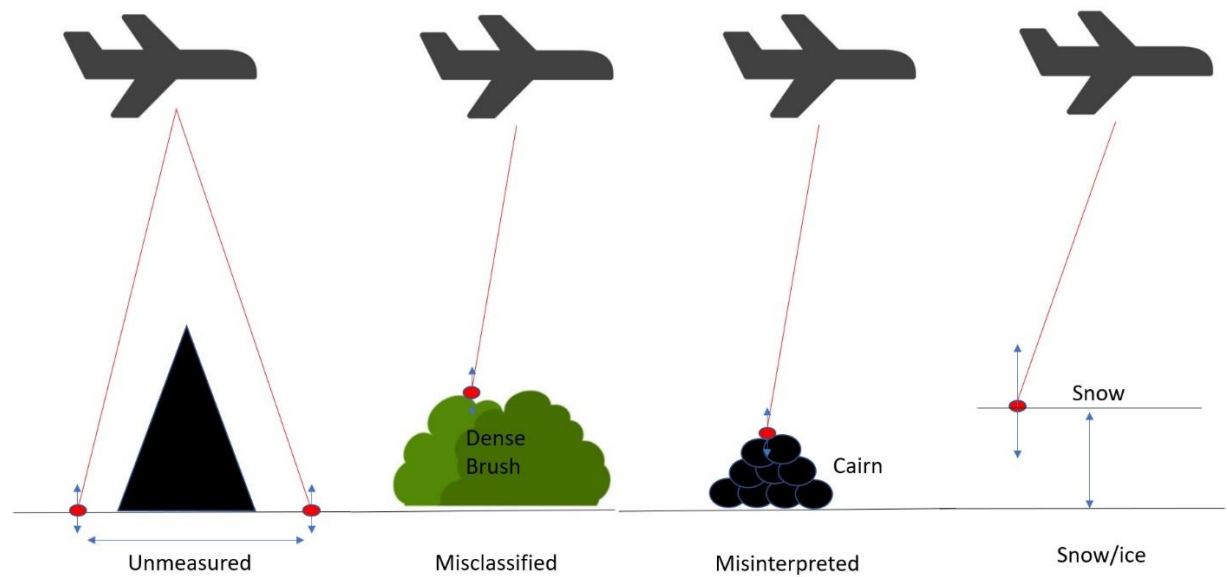

Fig. 1: The main error sources of LiDAR in mountainous terrain: sharp unmeasured features, dense brush misclassified as ground, human-made features misinterpreted as ground, and snow/ice that can melt over time. Note: these illustrations are schematic in nature.

**Methodology**

*Differential* GNSS *Ground Measurements*

To determine the accuracy of LiDAR data, we measured the elevations of 179 mountain summits spread across the continental US and southern Canada using dGNSS units. The authors are based in the US, thus field measurements were most practical in the US and southern Canada. All summits had also been measured with LiDAR surveys, and we compared the LiDAR-identified and measured summit location and elevation to the dGNSS – measured summit location and elevation. LiDAR surveys were conducted by the US Geological Survey in the US and the Canadian Geological Survey in Canada.

This set of peaks covered a broad range of mountain types including sharp rocky summits, summits with dense brush, above-treeline and below-treeline summits, summits with permanent icecaps, and summits with artificial human-made structures near the tops. We surveyed summits in 12 US states covering the Cascades, Rockies, and Appalachian ranges, and in two Canadian provinces.

While dGNSS measurements are not practical for all peaks, measuring a subset of peaks with dGNSS allows for more accurate error bounds to be assigned to LiDAR measurements for a large set of peaks.

We conducted static ground surveys with differential GNSS units on summits in Washington, Idaho, Montana, California, Utah, Nevada, Arizona, New Mexico, Michigan, Wisconsin, Kentucky, Pennsylvania in the continental US, and British Columbia and Nova Scotia in southern Canada (Figure 2). Summits surveyed included tree- and brush-covered summits (26), open rock summits (149), and icecapped summits (4). Each survey used at least one of a Trimble Promark 220 with Ashtech Antenna, a Trimble DA2, or an Emlid Reach 2.

For each survey, a 5x 10-arcminute Abney level was first used to determine the highest point on the mountain. To do this, we went to the coordinates of the highest point identified by LiDAR, then took Abney level measurements in all directions from that location. Then, the dGNSS unit was mounted on a tripod either using a 2.0m or 0.3m antenna rod or a flexible-leg tripod, and static survey data was collected. In the case of summits with vegetation, generally the 2.0m antenna rod was used to raise the receiver off the ground to reduce errors. For taller rocky summits where weight considerations were necessary, a 0.3m antenna rod was used.

For technical summits where lightweight equipment was even more important, a smaller flexible-leg tripod was used to mount the dGNSS exactly on the highest rock (Figure 2, Old Guard Peak). The distance from the antenna to the summit was then measured with a tape measure.

For summits with permanent icecaps, we measured the summit elevation in late summer at the approximate minimum snow-depth time of year. This is consistent with measurements of other icecapped peaks around the world such as Kebnekaise, Sweden (Holmlund and Holmlund 2019) and Mt. Blanc, France/Italy (Berthier et al. 2023).

There have historically existed just five icecap peaks in the entire contiguous US (Gilbertson 2025) and we measured four of these for this study, which is a small but significant dataset. We took dGNSS measurements within 1-2 years of when LiDAR data was collected, so changes due to potential melting would be minimal. We measured non-icecapped summits at a time of year when all snow had melted off. All measurements were taken between 2022 – 2025.

We processed static survey data from the US with the Online Positioning User Service (OPUS; https://geodesy.noaa.gov/OPUS/) and data from Canada with Canadian Spatial Reference System Precise Point Positioning (CSRS-PPP; https://webapp.csrs-scrs.nrcan-rncan.gc.ca/geod/tools-outils/ppp.php ). Measurements were generally taken for at least one hour for open rocky or ice summits, and for up to five hours for summits with vegetation. In all cases 95% confidence interval vertical errors were +/-0.03m or better.

To ensure the accuracy of the dGNSS units, we occupied USGS benchmarks with known elevations for comparison. We used the Promark and DA2 to occupy the McClure Rock (WA, USA), Curwood (MI, USA), Arvon (MI, USA), and Timms (WI, USA) benchmarks. For all benchmarks, measured elevations were within +/-0.03m of published USGS elevations.

To ensure the different dGNSS models gave consistent results, we resurveyed some peaks using alternate units and compared the results. We surveyed He Devil and She Devil peaks in Idaho with both the Emlid Reach 2 and the Promark. The resulting elevation measurements were within +/-0.03m. Similarly, we surveyed Big Craggy West and Big Craggy East in Washington with both the Promark and the DA2. The resulting elevation measurements were within +/-0.03m.

For peaks in the US, we derived elevations in NAVD88 vertical datum and NAD83 horizontal datum. These are the most updated and accurate vertical and horizontal datums in the US. For peaks in Canada we derived elevations in CGVD2013 vertical datum and NAD83 horizontal datum. These are the most updated and accurate vertical and horizontal datums in Canada.

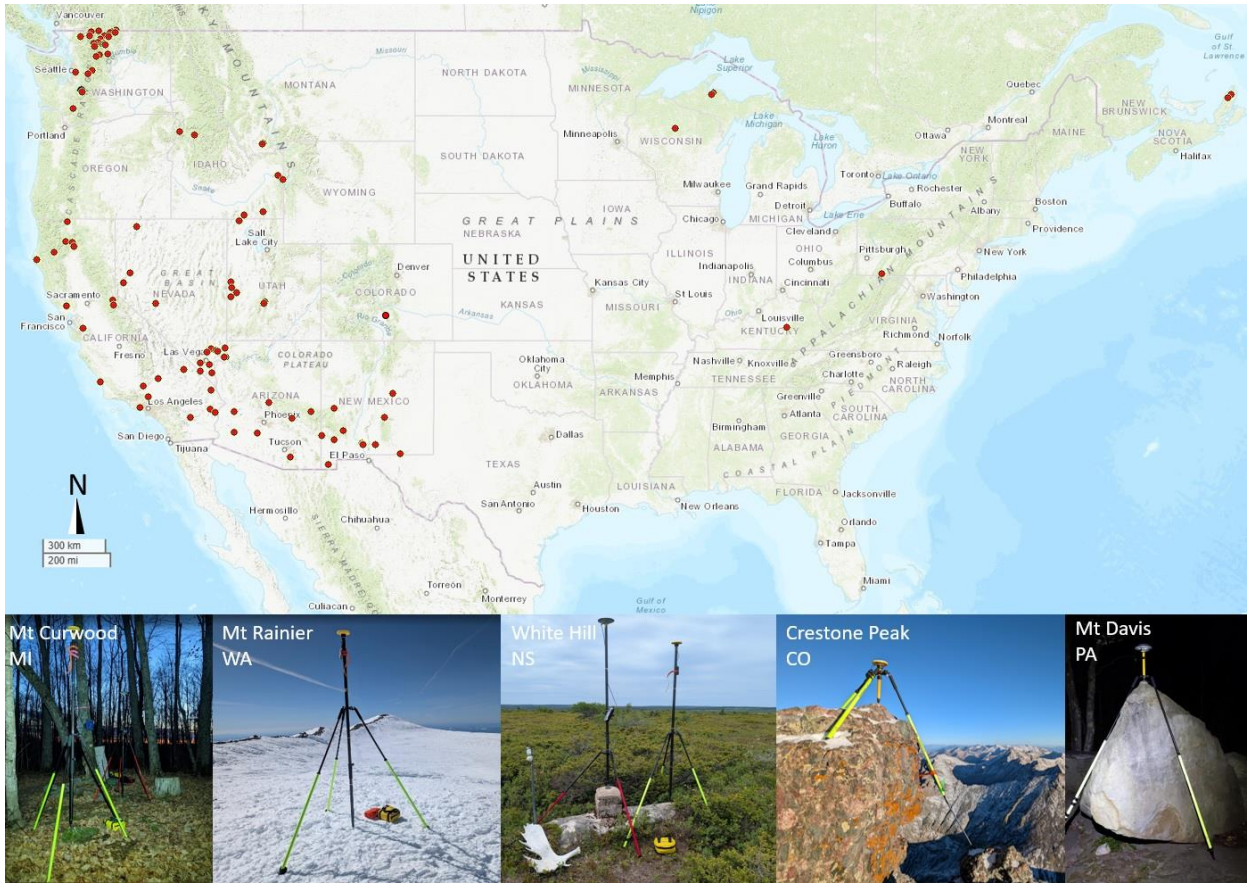

Fig 2: Locations of surveyed summits in the US and southern Canada. Photos of dGNSS units set up on Mt. Curwood, MI, Mt. Rainier, WA, White Hill, NS, Crestone Peak, CO, and Mt. Davis, PA.

*LiDAR Processing*

To process LiDAR data, we used three different automated methods and one manual method. For manual processing, we downloaded LiDAR raw point cloud data from the national map downloader app (https://apps.nationalmap.gov/downloader/ ), then opened the data using QGIS 3.38.2. We then displayed only points classified as ground or unclassified, and located the highest point on the mountain, throwing out obvious outlier points or noise. Washington, Pennsylvania, Michigan, and Wisconsin peaks were processed manually.

All LiDAR data collected in the US adhered to the USGS Lidar Base Specifications (NGP 2025). All data we used was QL2 or better. LiDAR data collected in Canada adhered to the National Standard of Canada Airborne LiDAR Data Acquisition Standards (CSA 2025).

We also used three automated processing methods. For each method we considered all samples in the point cloud within a 20m radius of the dGNSS-located summit latitude and longitude. For the highest return method, we algorithmically found the location of the highest sampled point of all points within this 20m radius and counted this as the summit. This could include points classified as ground, vegetation, snow, or unclassified.

For the second automated method, we first kept only points classified as ground in the original source data, and algorithmically located the highest ground point within this 20m radius.

Finally, for a third method, we reclassified the point cloud using the proprietary algorithm in the Lastools software package version 240605 (Hug et al. 2004) and located the highest such sampled ground point within the 20m radius. The Lastools algorithm is described in detail in section 3.3 of the referenced paper.

The Lastools algorithm is less conservative than the algorithm originally used to classify the point cloud, in the sense that it correctly classifies some points as ground that the original algorithm leaves unclassified. For example, in a sample of 2324 known peaks in USGS Lidar data collections in the state of Oregon, the Lastools classifier returned a highest ground sample with greater elevation than the original USGS classifier in 2219 cases (99.3%).

However, the Lastools algorithm also exhibits more false positives, often classifying buildings as ground, and sometimes classifying vegetation as ground. For each method we recorded the latitude, longitude, and elevation of the highest point.

Note that only summits in Washington, Pennsylvania, Michigan, Wisconsin, and Kentucky were analyzed using the manual method. All summits were analyzed using the three automated methods.

**Results**

*Vertical Errors*

We compared LiDAR summit elevations to dGNSS summit elevations for all peaks and calculated mean and one-sigma errors, assuming normal distributions. We categorized peaks as either permanent ice summit, forested summit, or open rocky summit. Mean and sigma values for vertical errors are shown in Table 2 for all summits for each of the four processing methods.

The manual processing method resulted in the lowest vertical errors (mean 0.03m, sigma 0.50m). The mean near 0m indicates that roughly as many errors were positive as negative. The manual processing method allowed for removing noise or erroneous measurements that might be picked up in the algorithmic methods.

The highest ground return method was the second most accurate (mean -0.31m; sigma 1.09m). The negative mean value is likely related to sharp summits getting missed by LiDAR and the highest ground being lower than the summit. One outlier with vertical error -12.9m significantly affected the results. Upon manual inspection, the data point closest to the sharp summit spire of Snoqualmie East Peak (Washington, USA) was left unclassified in the original dataset, so the highest ground return was much lower than the summit. This is a systemic classification error, which the highest ground return method is subject to.

Lastools was the next most accurate, with a positive mean and higher sigma value, 1.96m. Lastools had higher errors than the highest ground return method likely due to four outlier measurements with Lastools errors 7-14m.

The highest return method was the least accurate, likely because summits with trees or brush will generally yield positive error, as much as the height of the tree or brush.

Table 2: Vertical errors in LiDAR for all summits for each processing method, assuming normal distributions.

| Processing Method | Mean Error (m) | Sigma (m) |
| --- | --- | --- |
| Manual | 0.03 | 0.50 |
| Highest Ground Return | -0.31 | 1.09 |
| Highest Return | 4.64 | 9.83 |
| Lastools | 0.53 | 1.96 |

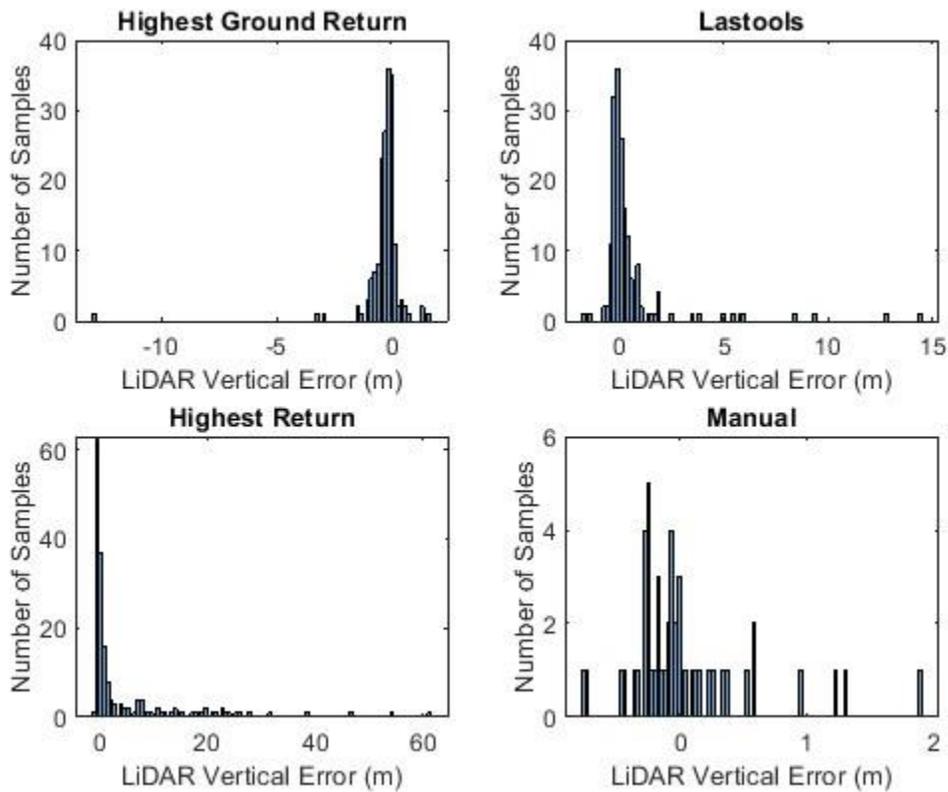

Fig 3: Histograms of LiDAR vertical error for all summits for all four different processing methods.

Table 3 shows the mean vertical errors and sigmas for forested summits. Mean errors are all positive, indicating brush and trees may be misclassified or misinterpreted as ground for all methods. This results in positive error. The highest return method resulted in the highest errors, because this method is susceptible to considering trees and vegetation as ground.

The remaining methods, in order of smallest to largest errors, where highest ground return, manual, and Lastools.

Table 3: Vertical error in LiDAR on tree-covered summits for all four processing methods. Additional columns show the number of measurements within a given error range for each method.

| Processing Method | Mean Error (m) | Sigma (m) | <-10m | -10m to -2m | -2m to -1m | -1m to 0m | 0m to 1m | 1m to 2m | 2m to 10m | >10m |
|---|---|---|---|---|---|---|---|---|---|---|
| Manual | 0.29 | 0.73 | 0 | 0 | 0 | 2 | 3 | 2 | 0 | 0 |
| Highest Ground Return | 0.14 | 0.66 | 0 | 0 | 1 | 8 | 14 | 3 | 0 | 0 |
| Highest Return | 14.15 | 12.03 | 0 | 0 | 0 | 0 | 3 | 2 | 8 | 13 |
| Lastools | 2.69 | 4.17 | 0 | 0 | 1 | 1 | 14 | 3 | 5 | 2 |

Table 4 shows the vertical error in LiDAR for ice-covered summits for all four processing methods. In this case the highest ground return method filtered out all results within a 20m radius of the summit because this region was all ice/snow and not identified as ground. Thus, no results are presented for this method.

All errors are positive, which could be a result of the icecap summits melting down after the LiDAR measurement. Thus, they were measured shorter with the dGNSS measurement, resulting in positive LiDAR error.

The remaining methods, in order of smallest to largest errors, where highest return, Lastools, and manual. Though, with the small dataset, it could be concluded all three of these methods performed similarly.

Table 4: Vertical error in LiDAR on ice-covered summits for all four processing methods. Additional columns show the number of measurements within a given error range for each method.

| Processing Method | Mean Error (m) | Sigma (m) | <-10m | -10m to -2m | -2m to -1m | -1m to 0m | 0m to 1m | 1m to 2m | 2m to 10m | >10m |
|---|---|---|---|---|---|---|---|---|---|---|
| Manual | 0.95 | 0.68 | 0 | 0 | 0 | 0 | 3 | 1 | 0 | 0 |
| Highest Ground Return | -- | -- | -- | -- | -- | -- | -- | -- | -- | -- |
| Highest Return | 0.55 | 0.28 | 0 | 0 | 0 | 0 | 4 | 0 | 0 | 0 |
| Lastools | 0.49 | 0.35 | 0 | 0 | 0 | 0 | 4 | 0 | 0 | 0 |

Table 5 shows the vertical error for just open rocky summits for all four processing methods. The manual method had the lowest errors, and the negative mean suggests sharp summits were often missed by LiDAR, resulting in undermeasurements.

The Lastools method had the next lowest error. It is possible misclassification of ground points contributed to these error numbers.

The highest ground return method had slightly higher error, and this was affected by the Snoqualmie East measurement mentioned previously. Once again, the negative mean indicates sharp summits were missed by the LiDAR surveys, resulting in undermeasurements.

The highest return had significantly higher error. This is likely because noise in the data was picked up in this method, while it was likely filtered out by the Lastools and highest ground return methods.

Table 5: Vertical error in LiDAR on open rocky summits for all four processing methods. Additional columns show the number of measurements within a given error range for each method.

| Processing Method | Mean Error (m) | Sigma (m) | <-10m | -10m to -2m | -2m to -1m | -1m to 0m | 0m to 1m | 1m to 2m | 2m to 10m | >10m |
|---|---|---|---|---|---|---|---|---|---|---|
| Manual | -0.11 | 0.26 | 0 | 0 | 0 | 30 | 8 | 0 | 0 | 0 |
| Highest Ground Return | -0.39 | 1.14 | 1 | 2 | 3 | 125 | 18 | 0 | 0 | 0 |
| Highest Return | 3.08 | 8.53 | 0 | 0 | 0 | 26 | 86 | 11 | 11 | 15 |
| Lastools | 0.15 | 0.84 | 0 | 0 | 1 | 84 | 55 | 5 | 4 | 0 |

Histograms are plotted for all four summit categories showing all four processing methods (Figure 4). These plots assume normal distributions for all data. Note that in some cases data is not well-modeled by normal distribution curves, such as for the highest return method for open rock and tree-covered summits as well as the Lastools method for tree-covered summits. A relatively small number of high-error peaks contribute to high sigma values, while Figure 3 and Tables 2-4 show there are in fact many peaks with low errors.

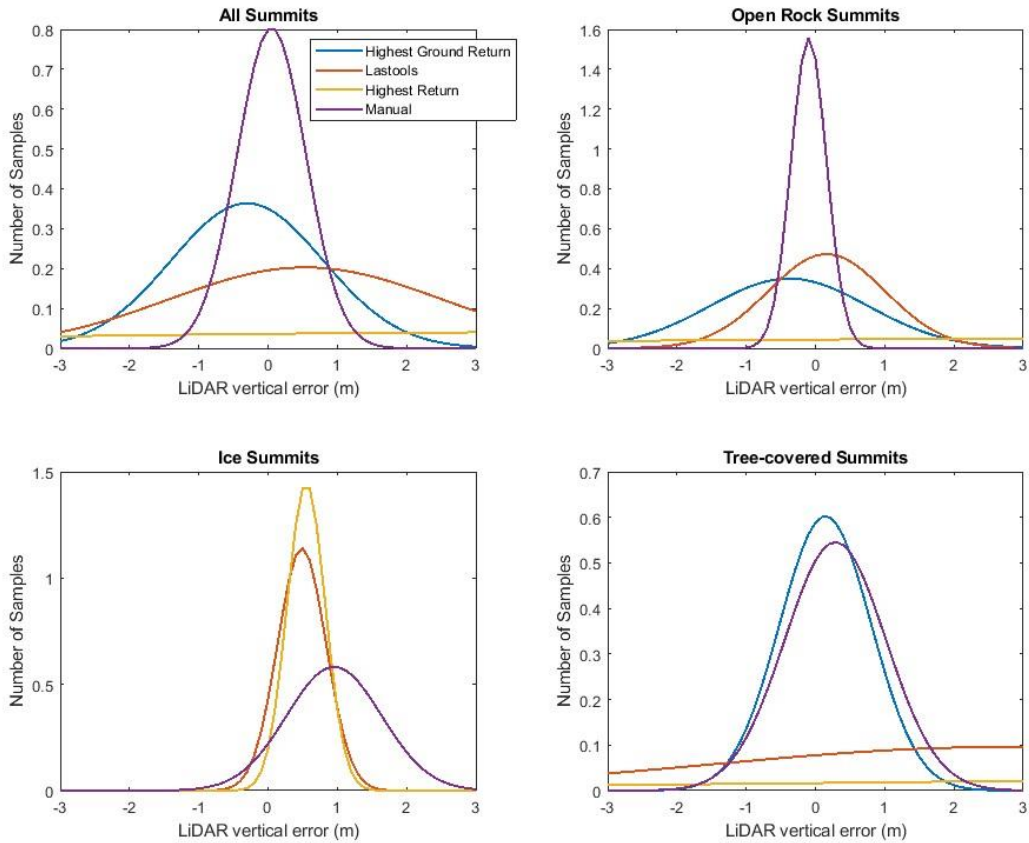

Fig 4: Normalized histograms of vertical LiDAR error for all four processing methods for all summits (upper left), open rock summits (upper right), ice-covered summits (lower left), and tree-covered summits (lower right).

Horizontal Error

For all summits we also calculated the horizontal error between the dGNSS summit coordinates and the LiDAR-identified summit coordinates for each processing method. To find the horizontal distance between coordinates we used the spherical law of cosines,

$$d = R\cos^{-1}(\sin(y_1)\sin(y_2) + \cos(y_1)\cos(y_2)\cos(x_2 - x_1)) \tag{1}$$

where $x_1$ and $x_2$ are the longitudes of the points of interest and $y_1$ and $y_2$ are the latitudes of the points of interest.

Figure 5 shows histograms of horizontal error for each of the four processing methods. Table 6 shows mean and sigma values for all measurements. In general, all mean values are positive, consistent with horizontal distance always being positive by definition. (For instance, an error north and an error south are both positive horizontal distances).

The highest ground return method has the smallest errors, with mean error values between 1-2m. This is consistent with the horizontal spacing of measured LiDAR points at 1-2m; that is, it is the inherent design error of LiDAR surveys.

The manual method has slightly higher error, followed by Lastools. The highest return has the highest error. This is likely because this data includes tree-covered summits, and the highest return may pick up a tree or brush that is a distance away from the summit.

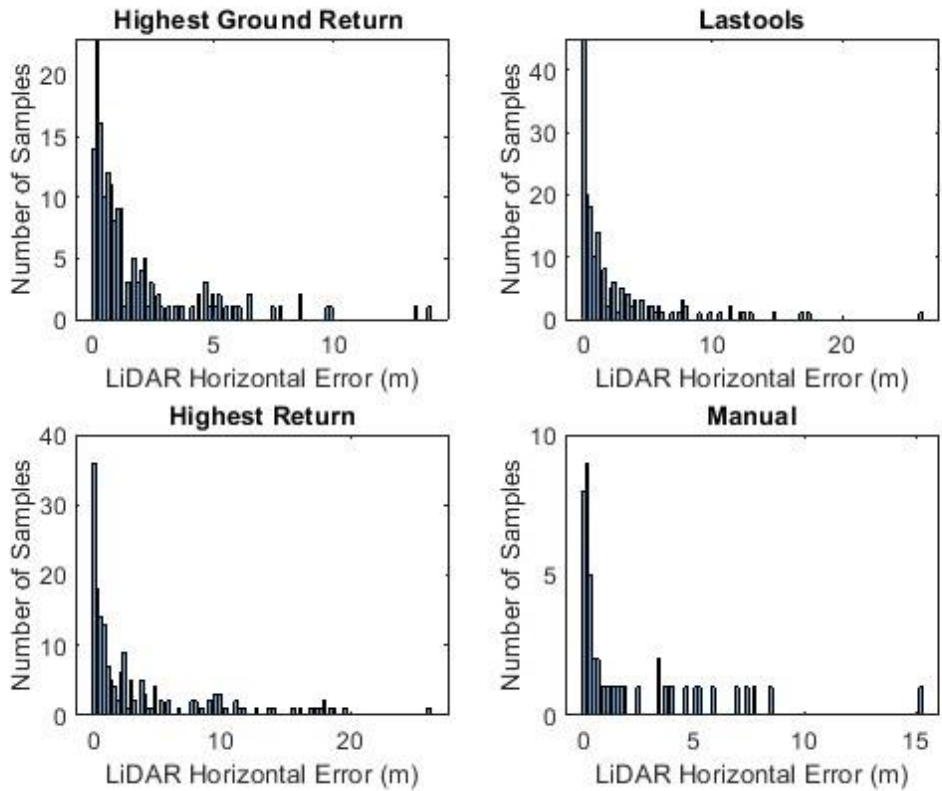

Fig 5: Histogram of horizontal error between LiDAR-identified summit and dGNSS summit for all summits for all four processing methods.

Table 6: Horizontal error for all summits for all four processing methods.

| Processing Method | Mean Error (m) | Sigma (m) |
| --- | --- | --- |
| Manual | 2.24 | 3.07 |
| Highest Ground Return | 1.78 | 2.41 |
| Highest Return | 3.78 | 5.17 |
| Lastools | 2.44 | 3.78 |

For tree/vegetation-covered summits, the highest ground return method has the lowest errors, and the highest return method has the highest errors. This is because all summits contain trees, and highest return method will not distinguish between trees and ground, leading to higher error (Table 7).

Table 7: Horizontal error for treed summits for all four processing methods. Additional columns show the number of measurements within a given error range for each method.

| Processing Method | Mean Error (m) | Sigma (m) | 0m to 1m | 1m to 2m | 2m to 10m | >10m |
|---|---|---|---|---|---|---|
| Manual | 5.93 | 4.63 | 1 | 0 | 5 | 1 |
| Highest Ground Return | 3.73 | 3.85 | 8 | 6 | 11 | 1 |
| Highest Return | 6.20 | 4.89 | 4 | 0 | 18 | 4 |
| Lastools | 5.58 | 4.78 | 3 | 4 | 15 | 4 |

Table 8 shows the horizontal error mean and sigma values for ice summits. As in the case of vertical errors for ice summits, the highest ground return method was not able to find any ground returns within 20m of the summit, thus no data is presented for this method.

The manual method has the lowest errors, and Lastools and the highest return method have higher errors, similar to each other. In all cases the ice summits melted after the LiDAR measurement, resulting in a different current summit elevation and different current summit location. This is why mean errors are high.

In particular, the ice-capped summit of Eldorado Peak (Washington, USA) melted significantly between the LiDAR measurement and the dGNSS measurement. Between 2022-2024 it melted 2.0m, which also resulted in the emergence of a new area of the icecap as the highest point on the mountain (moved location horizontally by over 20m). This peak thus contributed to the high horizontal errors in ice summit data. The summit area of Eldorado is a long narrow ice fin, nearly flat on top. The manual method identified a point on one end of the fin while the Lastools and highest return identified a point of similar but slightly lower elevation on the other side of the fin. This is why the manual method had much lower error in this case.

Table 8: Horizontal error for ice summits for all four processing methods. Additional columns show the number of measurements within a given error range for each method.

| Processing Method | Mean Error (m) | Sigma (m) | 0m to 1m | 1m to 2m | 2m to 10m | >10m |
|---|---|---|---|---|---|---|
| Manual | 4.78 | 2.09 | 0 | 0 | 4 | 0 |
| Highest Ground Return | -- | -- | -- | -- | -- | -- |
| Highest Return | 9.34 | 11.24 | 0 | 0 | 3 | 1 |
| Lastools | 8.97 | 11.45 | 0 | 0 | 3 | 1 |

Table 9 shows the horizontal errors for open rocky summits. The manual, highest ground return, and Lastools methods all have mean errors between 1-2m, which is near the horizontal resolution of the LiDAR horizontal measurements. Lastools has slightly higher mean and sigma values.

The highest return method has a much higher mean error and sigma value. This is likely due to noise in the data being misinterpreted as the summit. The noise is likely removed in the other three processing methods.

Table 9: Horizontal error for open rocky summits for all four processing methods.

| Processing Method | Mean Error (m) | Sigma (m) | 0m to 1m | 1m to 2m | 2m to 10m | >10m |
|---|---|---|---|---|---|---|
| Manual | 1.25 | 1.97 | 27 | 6 | 5 | 0 |
| Highest Ground Return | 1.42 | 1.88 | 92 | 26 | 32 | 1 |
| Highest Return | 3.22 | 4.83 | 77 | 17 | 38 | 15 |
| Lastools | 1.73 | 2.72 | 85 | 28 | 31 | 5 |

Figure 6 shows normalized histograms for all summits, open rocky summits, treed summits, and ice summits are shown for all four processing methods. These plots assume normal distributions of data, and plots are only shown for positive error values because horizontal error is defined to be positive. Note that in the ice summits plot the curves for Lastools and the highest return are affected by the Eldorado ice summit data as mentioned previously.

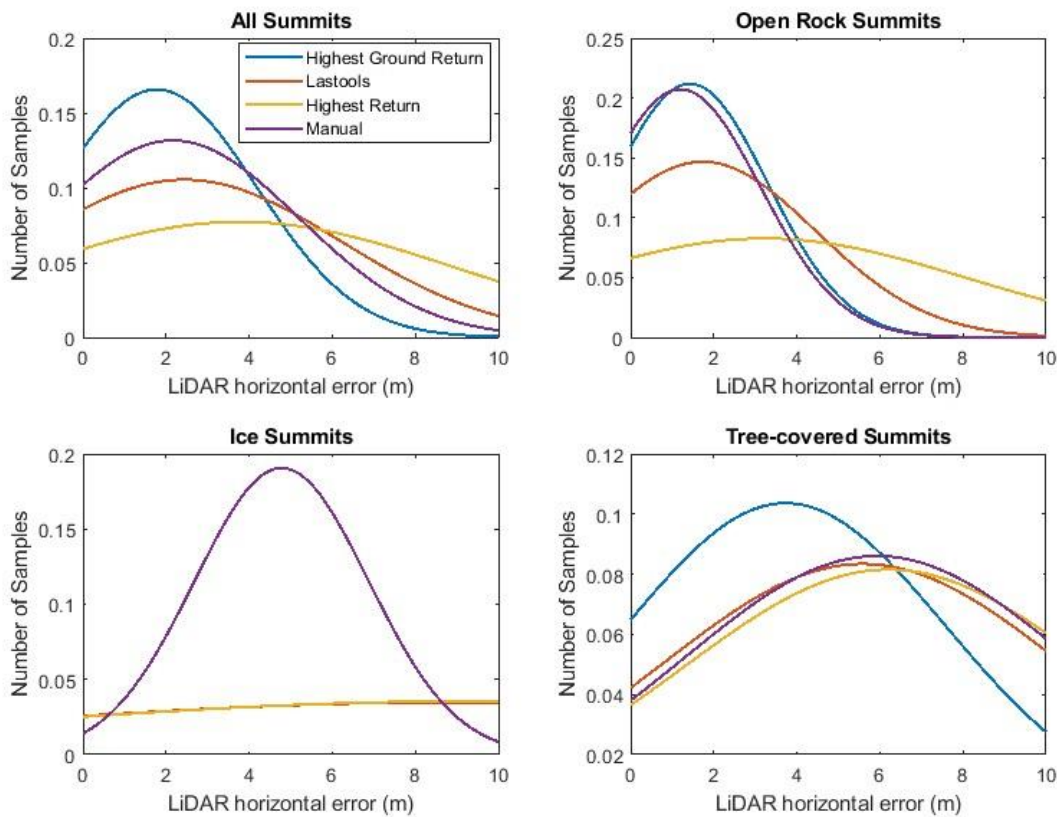

Fig 6: Normalized histograms of horizontal Lidar error for all four processing methods for all summits (upper left), open rock summits (upper right), ice-covered summits (lower left), and tree-covered summits (lower right).

Horizontal and Vertical Errors

Horizontal and vertical errors can be correlated. Thus, we present scatter plots of horizontal and vertical errors for all summits for all four processing methods (Figure 7).

In general, for the highest ground return, Lastools, and highest return methods, low vertical error is correlated with low horizontal error. However, for all three methods there are points measured with low vertical error but high horizontal error. There are fewer instances of high vertical error with low horizontal error.

The highest return method has only positive or zero vertical error. This is likely because all data sets either have trees or ice higher than the current summit, or noisy signals higher than the summit.

In general, for the manual method, there is minimal correlation between vertical and horizontal errors. A few cases had high horizontal errors with low vertical errors. These may have been cases of tree-covered summits where the summit location was misidentified when trees were misclassified as ground, or when an ice summit like Eldorado melted down and changed location. This could occur in the case of a relatively flat summit, with similar elevations extending across a large horizontal distance.

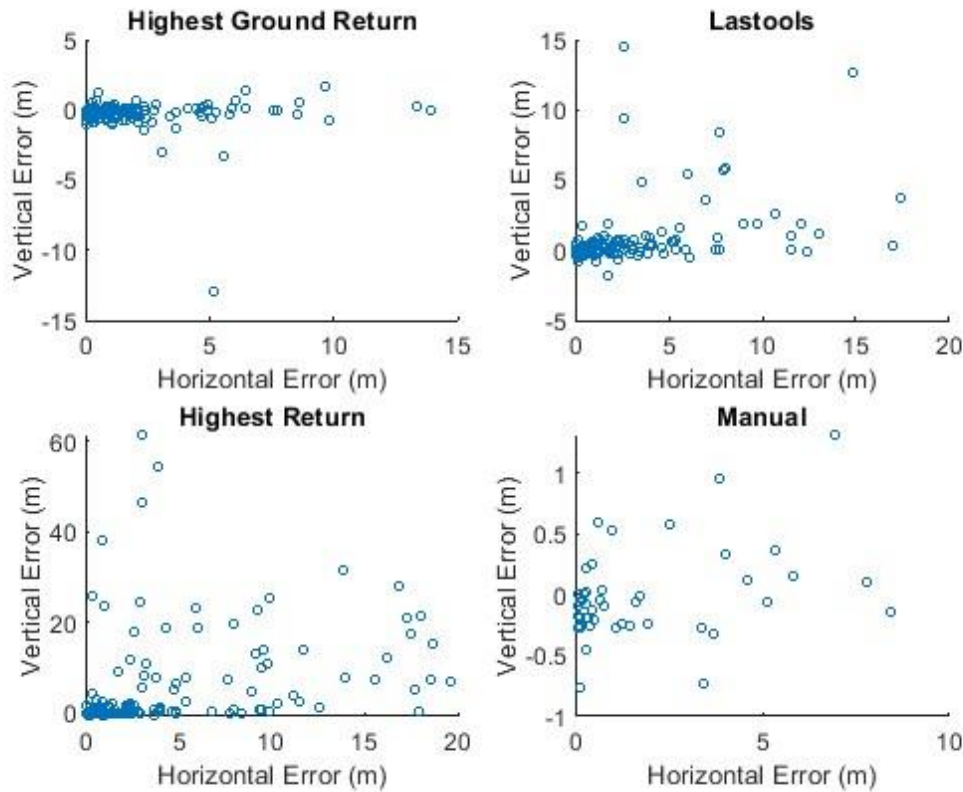

Fig 7: Plots of vertical error vs distance between Lidar summit point and dGNSS summit point for all four Lidar processing methods.

Special cases

We present several special cases where LiDAR errors were significant enough to misidentify state or provincial highpoints. Other cases resulted in significant errors for peaks with permanent icecaps at summits or summits that eroded over time.

*Michigan, USA – Mt. Curwood and Mt. Arvon*

The highest peak in the state of Michigan was identified by LiDAR measurements as Mt. Arvon. However, ground surveys with dGNSS measured a nearby peak, Mt. Curwood (Fig 1), as 0.12m taller (Mt. Curwood was measured at 603.29m and Mt. Arvon at 603.17m NAVD88 vertical datum). In this case, the LiDAR measurement missed the very top of the sharp summit of Mt. Curwood and underestimated its height. The LiDAR measurement was 0.30m horizontal away from the summit, resulting in a vertical undermeasurement of 0.13m.

On Mt. Arvon, LiDAR picked up an artificial human structure 9.00m away horizontally from the summit, resulting in a vertical overmeasurement of 0.12m. Thus, the dGNSS measurements showed Mt. Curwood is in fact the true state highpoint of Michigan.

*Pennsylvania, USA – Mt. Davis*

In Pennsylvania, LiDAR measured both the Mt. Davis North Summit and the Mt. Davis South Summit at 1.0m taller than Mt. Davis, the traditional state highpoint of Pennsylvania. However, ground surveys by dGNSS showed these measurements were in error. The north and south summits were covered in dense brush, and this was misclassified as ground in all four LiDAR processing methods, giving a vertical error of 1.3m over-measurement in the LiDAR measurements.

Mt. Davis itself (Fig 1) was not covered in dense brush but had a very sharp summit that LiDAR measurements missed, resulting in a vertical error of 1.0m under-measurement. The dGNSS measurements showed that Mt. Davis is in fact the state highpoint of Pennsylvania.

*Nova Scotia, Canada – Western Barren and White Hill*

In Nova Scotia, LiDAR identified a new provincial highpoint, Western Barren, different than the previously-accepted White Hill. However, both peaks were covered in dense brush which contributed to significant errors in LiDAR.

For each peak, all three automated LiDAR processing methods misidentified the highest location and misclassified dense brush as ground. For Western Barren, Lastools identified a bush 20m horizontal from the highest ground as the summit, and the highest return was a different bush 10m horizontal from the highest ground. The highest ground return was 3m from the highest ground, but was also dense brush. Each of these returns was in error of 0.7m – 1.4m vertical from the highest ground.

For White Hill (Fig 1), all three automated LiDAR measurements also misclassified dense brush as ground. Lastools and the highest return method identified a bush 16m horizontal from the highest ground. The highest ground return identified a different bush also 16m horizontal from the highest ground. Vertical errors were 0.6-1.2m.

Ground measurements with dGNSS measured Western Barren is 1.59m taller than White Hill (Western Barren 531.32m, White Hill 529.73m CGVD2013 vertical datum). Thus, LiDAR correctly identified the provincial highpoint, but vertical errors in LiDAR measurements were comparable to the height difference between the peaks, meaning the LiDAR measurements alone were not definitive for which peak is higher.

*Washington – Mt. Rainier*

In several instances, mountain elevations changed over time, and the LiDAR measurements were outdated. Mt. Rainer (Fig 1), the highest peak in Washington state, was measured by LiDAR in 2007, showing the summit was Columbia Crest, a permanent icecap. A subsequent LiDAR measurement in 2022 (published in 2024) showed Columbia Crest short enough to no longer be the highest point on the mountain.

DGNSS measurements in 2024 showed Columbia Crest had melted down 6.0m since 2007 and 0.9m since 2022 and was no longer the highest point on the mountain. The new summit location was measured to be a rocky point on the southwest crater rim, 133m away horizontal. The 2022 Lidar measurement measured the southwest rim summit 0.3m shorter than the 2007 Lidar measurement, which is likely explained by the sharpness of the rocky summit. The 2022 Lidar measurement missed the top of the rock, while the 2007 measurement hit closer to the top of the summit and agreed within 0.03m of the dGNSS measurement.

Other icecap peaks in Washington – Colfax, Liberty Cap, East Fury, and Eldorado Peak – were measured to have lost up to 2.0m of elevation since the most recent LiDAR surveys in 2022. This loss was most likely due to melting (Gilbertson et al., 2025).

*Washington, USA – Mt. St Helens*

Using dGNSS, one peak, Mt. St Helens, was found to have lost significant elevation due to erosion compared to the most recent LiDAR measurements. The summit was last measured by LiDAR in 2018 but has eroded approximately 0.3m between then and 2024, and 3.7m between 1989 and 2024. We found based on previous LiDAR measurements and multiple dGNSS measurements that the loss rate is linear (0.10m/year since 1989 with an R squared value of 0.98; Fig 8).

Mt. St Helens was measured by triangulation in 1989, then by LiDAR in 2003, 2009, 2017, and 2018. Each measurement showed a lower summit elevation than the previous measurement, consistent with a hypothesis of summit erosion. The summit is located on the south side of the crater rim and has a gentle southern slope away from the crater and a steep northern slope leading to the inner crater. We hypothesize that rocks erode from the summit down the north face every year, leading to shrinking summit elevations. It is possible that cornice collapse in the spring contributes to this erosion.

The accuracy of LiDAR measurements from previous years is not known with certainty because of the changing nature of the summit. The dGNSS measurements highlight the fact that LiDAR measurements may be outdated for mountains that change elevation yearly.

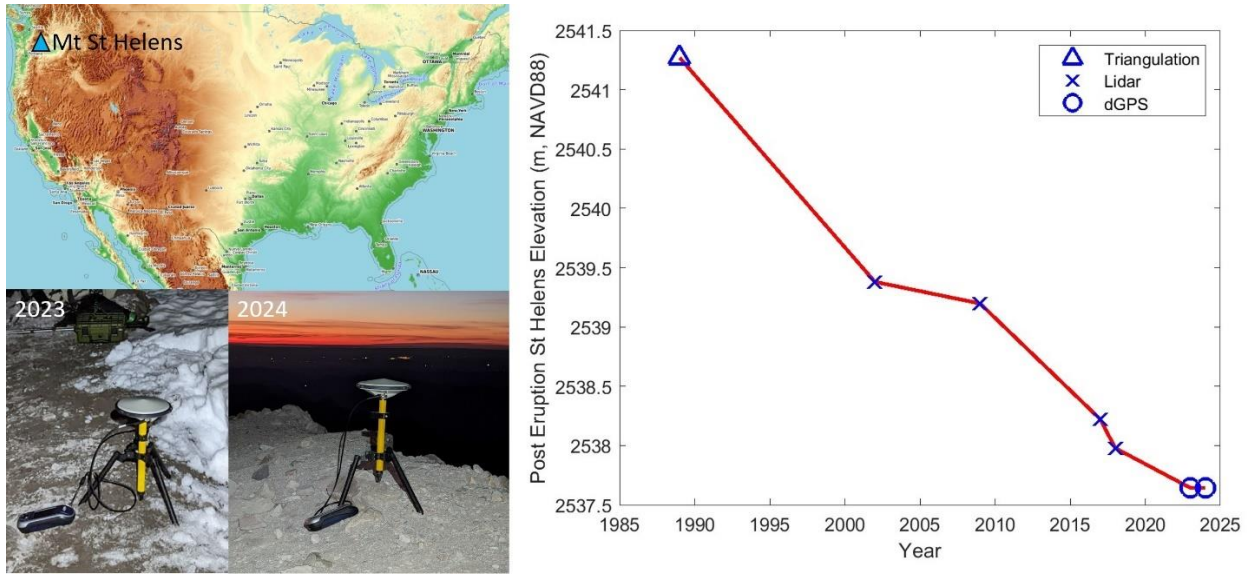

Fig 8: Elevation of Mt. St Helens since 1989.

*Colorado, USA – Crestone Peak and East Crestone*

In one instance, LiDAR was in error by 0.94m for a summit elevation because a measurement sampled what was possibly a person on a summit. The point was misclassified as ground, leading to an overmeasurement. This occurred on Crestone Peak in Colorado (Fig 1), an open rocky summit. The summit itself is relatively flat with no rock cairns, but LiDAR point cloud data indicated a point 0.94m higher than the surrounding points over a horizontal distance of just 0.30m. Ground surveys indicated this spike did not exist in reality. However, that height difference is a similar height to a sitting person. This is a very popular summit in Colorado and the LiDAR measurement was taken during the summer, so it would not be unusual for a person to be sitting on the summit then.

The LiDAR error resulted in Crestone Peak erroneously being considered a ranked 14er (14,000ft peak), thus making it a popular mountaineering objective in Colorado. Instead, nearby peak East Crestone is actually 0.09m taller based on dGNSS measurements, and this means East Crestone is in fact a ranked 14er and Crestone Peak is not.

**Discussion**

LiDAR is a valuable tool for finding the elevations and latitude/longitude coordinates of mountain summits. DEMs exist to approximate points between measured points in LiDAR data sets, but if a summit is not directly measured, there is no guarantee a model will accurately capture its location and elevation. It is difficult to know for certain if a summit is directly measured without comparing to ground surveys, and there is no guarantee a DEM will be more accurate at determining the location and elevation of a summit than by using the highest measured data point.

Thus, it is important to understand the accuracy of LiDAR point cloud data in determining summit locations and elevation. However, LiDAR data still has errors and should be used with caution, especially in topographically variable terrain. The four methods analyzed in this study for processing LiDAR data for mountain summits have tradeoffs in application and in accuracy.

The manual processing method in general provides the greatest accuracy of all methods, but also requires the most time to process data. One-sigma vertical errors, a good benchmark for comparison for vertical accuracy, range from 0.26m for open rocky summits to 0.73m for summits covered in vegetation. Mean horizontal errors, a benchmark for horizontal accuracy, range from 1.25m for open rocky summits to – 5.93m for tree-covered summits.

The Lastools and highest ground return methods are generally the next most accurate and require less time to process data than the manual method. Lastools has one-sigma vertical errors ranging from 0.35m for ice-capped summits to 4.17m for tree-covered summits. The highest ground return method has one-sigma vertical errors ranging from 0.66m on tree-covered summits to 1.14m on open rocky summits. Mean horizontal errors range from 1.42m to 8.97m for both methods. These errors are generally higher than the manual processing method.

One disadvantage of the highest ground return method is that it cannot correctly identify the summits of icecapped peaks. The Lastools method can correctly identify these summits. The Lastools method also has significantly higher one-sigma vertical errors on tree-covered summits than the highest ground return method.

The highest return method is the least accurate in general for vertical and horizontal accuracy. Noisy signals and trees can be misidentified as the summit, leading to both horizontal and vertical error. One-sigma vertical errors can be up to 12.03m. Horizontal errors are less, but still range up to 11.24m.

The one previous study that compared static dGNSS measurements on mountain summits to LiDAR point cloud data on mountain summits (Maciuk 2021) found vertical errors comparable to those in our study for open rocky summits. Both results used manual processing. However, direct comparison is still difficult because of the small sample size of that study (six peaks).

The two previous studies in flat tree-covered terrain (Schmelz and Psuty 2019, Liu 2022), had smaller vertical errors than the results in this study. This is possibly because tree-covered mountains have the additional potential error source of sharp summits, which were not present in the flat tree-covered terrain. This would likely lead to higher vertical errors.

Unsurprisingly, LiDAR errors in flat parking lots (Elaksher and Alharthy 2023) were significantly lower than errors for any category of mountains studied. This is likely because none of the main error sources in mountains (sharp summits, vegetation, ice, human-made structures) generally exist in flat paved terrain.

Given the observed errors in LiDAR data for mountain summits, it is important to treat LiDAR-derived summit elevations with caution. In the cases of Pennsylvania and Michigan, state highpoints were misidentified with LiDAR data. This has implications for recreation and tourism in these states. Accuracy-minded mountain climbers should climb Mt. Curwood to reach the true state highpoint of Michigan and Mt. Davis to reach the true state highpoint of Pennsylvania.

In the case of icecapped peaks like Mt. Rainier, LiDAR-derived summit elevations can soon become outdated and have high vertical errors. Before the 2024 dGNSS measurements of Mt. Rainier, it was thought that Columbia Crest, the former recognized summit, was maintaining constant elevation and was still the highest point on the mountain. However, dGNSS measurements in 2024 showed that Columbia Crest had in fact melted 5.0m since the 2007 LiDAR measurement, and is no longer the highest point on Mt. Rainier. This measurement indicated that global warming is affecting the highest elevations of peaks in the continental US, and the result was not evident based on LiDAR data.

In this study we have shown that LiDAR measurements are prone to vertical and horizontal errors in identifying elevations and locations of mountain summits. These errors have implications for cartography, water resource management, tourism, and understanding the effects of global warming on peaks.

**Acknowledgements:**

Surveying equipment was provided by Seattle University and Trimble. The American Alpine Club provided funding for surveying icecap peaks in Washington.

**Declaration of Conflicting Interests**

The authors declare no potential conflicts of interest with respect to the research, authorship, and/or publication of this article.

**Funding**

The American Alpine Club provided funding for surveying icecap peaks in Washington.

**Ethical approval and informed consent statements**

This article does not contain any studies with human or animal participants.

**Data availability statement**

Raw measurement data is available as a supplemental file.

**Supplementary Materials**

Table S1: Peaks surveyed, with notes of summit type (R = open rocky, I = ice, T = tree-covered) and dGNSS-surveyed summit coordinates and elevation (m, NAVD88 vertical datum for US peaks, CVGD2013 for Canadian peaks).

| Peak | State/Province | Type | Lat | Long | Elevation (m) |
|---|---|---|---|---|---|
| Mount Rainier | WA | R | 46.851731 | -121.760395 | 4391.01 |
| Mount Rainier - Columbia Crest | WA | I | 46.852950 | -121.760571 | 4387.84 |
| Liberty Cap | WA | I | 46.862813 | -121.774638 | 4296.80 |
| Sherman Peak | WA | R | 48.767268 | -121.813697 | 3090.28 |
| Colfax Peak | WA | I | 48.771634 | -121.843953 | 2878.71 |
| Mount Maude | WA | R | 48.137695 | -120.803666 | 2768.32 |
| Eldorado Peak | WA | R | 48.53741 | -121.134501 | 2705.98 |
| Katsuk Peak | WA | R | 48.578449 | -120.888902 | 2646.61 |
| Big Craggy Peak | WA | R | 48.763498 | -120.329729 | 2584.19 |
| Blackcap Mountain | WA | R | 48.803566 | -120.542834 | 2562.48 |
| Solitude Peak | WA | R | 48.962964 | -121.24769 | 2562.61 |
| Big Snagtooth | WA | R | 48.534263 | -120.588409 | 2553.95 |
| Mount Ballard | WA | R | 48.686066 | -120.757928 | 2552.85 |
| Chalangin Peak | WA | R | 48.064408 | -120.993092 | 2551.39 |
| Mount Formidable | WA | R | 48.416278 | -121.066957 | 2547.34 |
| Castle Peak | WA | R | 48.981991 | -120.862054 | 2544.50 |
| Windy Peak | WA | R | 48.928205 | -119.970556 | 2542.06 |
| Mount Fury - East Peak | WA | R | 48.810753 | -121.312163 | 2537.83 |
| Mount Saint Helens | WA | R | 46.191444 | -122.195816 | 2537.64 |
| Switchback Mountain | WA | R | 48.175751 | -120.358334 | 2537.80 |
| Apex Mountain | WA | R | 48.961174 | -120.14479 | 2532.00 |
| Mount Ballard - North Peak | WA | R | 48.688703 | -120.759736 | 2531.33 |
| Andrew Peak | WA | R | 48.920147 | -120.227842 | 2529.44 |
| Ares Tower | WA | R | 48.560018 | -120.59543 | 2500.21 |
| Armstrong Mountain Northwest | BC | R | 49.001013 | -119.946089 | 2480.07 |

| Name | State | Type | Lat | Lon | Elev |
|---|---|---|---|---|---|
| Armstrong Mountain | WA | R | 48.998007 | -119.939113 | 2477.69 |
| South Mirror Image | WA | R | 48.514641 | -120.53221 | 2461.75 |
| Shelokum Mountain | WA | R | 48.511208 | -120.529828 | 2461.14 |
| Freezer | WA | R | 48.123046 | -120.804038 | 2448.50 |
| Wallaby Peak | WA | R | 48.506879 | -120.615243 | 2436.21 |
| Wolftit Peak | WA | R | 48.489652 | -120.517104 | 2435.32 |
| Mount Daniel | WA | R | 47.565031 | -121.180891 | 2431.39 |
| Mount Daniel - Middle Summit | WA | R | 47.565674 | -121.178209 | 2425.17 |
| Snoqualmie Mountain East | WA | R | 47.458779 | -121.414331 | 1915.85 |
| Snoqualmie Mountain | WA | T | 47.458869 | -121.416533 | 1914.94 |
| Mount Davis | PA | T | 39.785989 | -79.176762 | 979.38 |
| Mount Arvon | MI | T | 46.755828 | -88.155317 | 603.17 |
| Mount Curwood | MI | T | 46.703004 | -88.239504 | 603.29 |
| Pearson Hill | WI | T | 45.447742 | -90.179217 | 594.60 |
| Timms Hill | WI | T | 45.450741 | -90.195298 | 594.94 |
| Western Barren | NS | T | 46.597981 | -60.760575 | 531.32 |
| White Hill | NS | T | 46.702546 | -60.598578 | 529.73 |
| Cougar Mountain | WA | T | 47.519412 | -122.092756 | 491.89 |
| Cougar Mountain - Clay Pit Peak | WA | T | 47.526359 | -122.092109 | 472.32 |
| Carne Peak | WA | R | 48.088454 | -120.803294 | 2161.27 |
| Katsuk W | WA | R | 48.578432 | -120.889289 | 2646.52 |
| Big Craggy E | WA | R | 48.76308 | -120.328068 | 2584.12 |
| Davis North | WA | T | 39.798462 | -79.171326 | 978.26 |
| Davis South | WA | T | 39.783602 | -79.177715 | 978.26 |
| Bear Mtn SE | WA | T | 37.527063 | -84.251608 | 513.10 |
| Eldorado Ice | WA | I | 48.53759 | -121.13414 | 2703.48 |
| Sentinel Peak | WA | R | 48.356236 | -121.040838 | 2516.70 |
| Old Guard | WA | R | 48.355924 | -121.035845 | 2517.53 |
| Arica Mountains High Point | CA | R | 34.0258434 | -114.9339337 | 657.33 |
| Spot 645 (Arica Mountains) | CA | R | 34.0245482 | -114.9331692 | 656.14 |
| Arrow Canyon Range High Point | NV | R | 36.6633224 | -114.8873008 | 1593.37 |

| Name | State | Type | Latitude | Longitude | Elevation |
|---|---|---|---|---|---|
| South Arrow Canyon Peak | NV | R | 36.645209 | -114.8816758 | 1592.00 |
| Big Maria Mountains High Point | CA | R | 33.8679636 | -114.6690264 | 1031.72 |
| Big Maria North | CA | R | 33.8783122 | -114.6632861 | 1030.77 |
| False Maria | CA | R | 33.8604536 | -114.6712433 | 1030.89 |
| Bonanza King | CA | R | 41.0970231 | -122.6244629 | 2159.17 |
| Bonanza King South | CA | R | 41.0940937 | -122.6257339 | 2157.28 |
| Bull Mountain | UT | R | 41.9113276 | -113.3660117 | 3031.02 |
| Dunn Benchmark | UT | R | 41.904829 | -113.388814 | 3025.90 |
| California Ridge High Point | NV | R | 36.5714833 | -114.6048193 | 959.18 |
| Peak 3141 (California Ridge) | NV | R | 36.5781043 | -114.6032221 | 957.29 |
| Peak 3140 (California Ridge) | NV | R | 36.5766562 | -114.6034582 | 957.19 |
| Peak 3138 (California Ridge) | NV | R | 36.5806927 | -114.6030584 | 956.43 |
| Camp Wilderness Ridge | NM | T | 32.017928 | -104.799027 | 2264.54 |
| Camp Wilderness Ridge Peak 7416 | NM | R | 32.0215107 | -104.8082378 | 2260.49 |
| Camp wilderness Ridge 2nd Contour | NM | T | 32.022904 | -104.804154 | 2261.68 |
| Little Pilot Peak | NV | R | 38.5312895 | -117.8234449 | 2466.50 |
| Peak 8087 (Cedar) | NV | R | 38.5340156 | -117.837383 | 2464.89 |
| Corral Hollow Hill | CA | R | 38.4831365 | -120.0765841 | 2490.25 |
| Corral hollow hill southeast | CA | T | 38.477918 | -120.073779 | 2487.75 |
| Corral hollow hill southwest | CA | R | 38.477783 | -120.079615 | 2485.46 |
| Dona Ana Mountains High Point | NM | R | 32.4581768 | -106.7843549 | 1777.38 |
| Peak 5824 (Dona Ana) | NM | R | 32.4576444 | -106.7830213 | 1775.09 |
| Dona Ana Peak | NM | R | 32.4541404 | -106.788697 | 1775.09 |

| Name | State | Type | Latitude | Longitude | Elevation |
|---|---|---|---|---|---|
| Summerford Mountain | NM | R | 32.5134792 | -106.8206639 | 1775.09 |
| Empire Mountains High Point | AZ | R | 31.8852367 | -110.6388686 | 1703.13 |
| Flowery Range High Point | NV | R | 39.3924428 | -119.5224619 | 2238.48 |
| Peak 7344 (Flowery Range) | NV | R | 39.3957802 | -119.529451 | 2238.36 |
| Peak 7325 (Flowery Range) | NV | R | 39.3942665 | -119.5259433 | 2233.76 |
| Rattler Benchmark | NV | R | 39.4002992 | -119.5299189 | 2238.21 |
| Harcuvar Peak | AZ | R | 33.9184688 | -113.6375174 | 1408.60 |
| Harcuvar Benchmark | AZ | R | 33.9163349 | -113.6405166 | 1408.24 |
| Hayfork Bally | CA | R | 40.6595428 | -123.2119739 | 1914.14 |
| Hayfork Bally Benchmark Outcrop | CA | R | 40.6587774 | -123.2186908 | 1913.53 |
| Ireteba Peaks | NV | R | 35.6068838 | -114.8314811 | 1546.80 |
| Ireteba Peaks - South Peak | NV | R | 35.594607 | -114.8290952 | 1543.17 |
| Irish Hills High Point | CA | T | 35.2170328 | -120.775885 | 556.69 |
| Saddle Peak | CA | T | 35.2223967 | -120.79309 | 555.47 |
| Jarilla Mountain | NM | R | 32.4356546 | -106.118134 | 1617.97 |
| Spot 5295 (Jarilla Mountains) | NM | R | 32.4374577 | -106.1150664 | 1615.53 |
| Juniper Pk | NV | R | 39.7955803 | -119.1831808 | 2199.10 |
| Spot 7203 (Truckee Range) | NV | R | 39.8006838 | -119.1859669 | 2196.54 |
| Mesa Leon | NM | R | 34.723479 | -105.194491 | 1933.65 |
| Mesa Leon West Contour | NM | R | 34.7220899 | -105.2031249 | 1929.96 |
| Tyrone Benchmark | NM | R | 32.6620583 | -108.3317322 | 2024.45 |
| Spot 6627 (Little Burro) | NM | R | 32.6584106 | -108.3348933 | 2019.85 |
| Mangas Mountain | NM | R | 34.0519525 | -108.3070578 | 2953.54 |

| Name | State | Type | Latitude | Longitude | Elevation |
|---|---|---|---|---|---|
| Mangas Communication Site | NM | T | 34.041798 | -108.30402 | 2950.07 |
| Spot 9695 (Mangus Mountain) | NM | R | 34.042614 | -108.3036501 | 2950.52 |
| Maricopa Mountains High Point | AZ | R | 32.9475434 | -112.3930465 | 997.73 |
| Maricopa Mountains NW contour | AZ | R | 32.9484076 | -112.3939252 | 997.40 |
| McKnight Mountain | NM | R | 33.0518515 | -107.8502752 | 3099.48 |
| Mcknight Mountain North | NM | R | 33.0570911 | -107.8543662 | 3097.44 |
| Monroe Peak | UT | R | 38.5360987 | -112.0734356 | 3424.09 |
| Monroe Peak Middle Peak | UT | R | 38.5391942 | -112.0728031 | 3415.31 |
| Glenwood Mountain | UT | R | 38.6253405 | -112.0156913 | 3423.24 |
| Odell Benchmark | CA | R | 34.543371 | -118.234797 | 1592.52 |
| Peak 5218 (Sierra Pelona) | CA | R | 34.5436186 | -118.2330125 | 1591.36 |
| Peak 5214 Sierra Pelona) | CA | R | 34.5440765 | -118.228374 | 1589.11 |
| Peak 5211 (Sierra Pelona) | CA | R | 34.546749 | -118.216956 | 1588.19 |
| Overton Ridge | NV | R | 36.5319732 | -114.511157 | 702.26 |
| Overton Ridge North | NV | R | 36.534694 | -114.5104484 | 700.61 |
| Oxford Peak | ID | R | 42.2674825 | -112.0964238 | 2831.32 |
| Oxford 2 Benchmark | ID | R | 42.2696051 | -112.0978598 | 2830.25 |
| Palomas Mountains High Point | AZ | R | 32.993581 | -113.634534 | 579.58 |
| Peak 1900 (Palomas Mountains) | AZ | R | 33.0144073 | -113.6349922 | 579.03 |
| Rainbow Ridge South | CA | R | 40.3346444 | -124.1102755 | 1102.98 |
| Rainbow Ridge North | CA | T | 40.3532373 | -124.1265565 | 1098.86 |

| Name | State | Type | Latitude | Longitude | Elevation |
|---|---|---|---|---|---|
| Rand Benchmark | CA | R | 35.3391717 | -117.6808855 | 1444.45 |
| Government Peak | CA | R | 35.35172 | -117.6732461 | 1447.86 |
| Sacramento Mountains High Point | CA | R | 34.8635561 | -114.8825913 | 1012.70 |
| Bannock Benchmark | CA | R | 34.8367225 | -114.8584231 | 1010.87 |
| Salt Creek Mountain | CA | R | 40.8769918 | -122.1769918 | 1331.67 |
| Salt Creek Mountain West Contour | CA | T | 40.8773039 | -122.1798656 | 1325.36 |
| Hogback Mountain | MT | R | 44.8944965 | -112.1236156 | 3231.52 |
| Sunset Peak | MT | R | 44.8560176 | -112.1469119 | 3228.47 |
| State Line Hills High Point | NV | R | 35.6778724 | -115.4492973 | 1643.18 |
| Peak 5388 (State Line Hills) | NV | R | 35.6816253 | -115.4504554 | 1642.35 |
| Steamboat Benchmark | ID | R | 45.2124678 | -115.7551509 | 2603.97 |
| Steamboat Benchmark South Contour | ID | R | 45.2092794 | -115.7528255 | 2602.35 |
| Oliver Peak | ID | R | 43.5197124 | -111.0557127 | 2743.08 |
| South Oliver Peak | ID | R | 43.5084752 | -111.0577227 | 2747.01 |
| Thunder Mountain | CA | R | 38.6767108 | -120.0854734 | 2872.34 |
| Thunder Mountain - West Peak | CA | R | 38.6742333 | -120.0908463 | 2871.19 |
| Tombstone Mountain | CA | R | 41.0474365 | -122.254635 | 1710.17 |
| Tombstone Mountain South | CA | R | 41.0465465 | -122.2541791 | 1709.56 |
| Tucson Mountain | NM | T | 33.6440317 | -105.6360393 | 2532.64 |
| Vanderbilt Peak | NM | R | 32.8518111 | -108.9969619 | 2064.72 |
| Foote Range High Point | UT | R | 39.4438536 | -113.8073793 | 1924.99 |
| Foote Range Northeast | UT | R | 39.4502161 | -113.7993887 | 1919.17 |

| Name | State | Type | Latitude | Longitude | Elevation |
|---|---|---|---|---|---|
| Conger Range Hight Point | UT | R | 39.1979555 | -113.7925661 | 2106.26 |
| Conger Range North | UT | R | 39.2001976 | -113.7909591 | 2105.71 |
| Juniper Benchmark [offset] | UT | T | 38.8311329 | -113.8097975 | 2383.63 |
| Juniper Benchmark [actual] | UT | T | 38.8320521 | -113.8099183 | 2383.17 |
| King Top | UT | R | 38.980899 | -113.5304405 | 2545.78 |
| Peak 8352 | UT | R | 38.9992399 | -113.5001886 | 2545.60 |
| Pine Mountain | AZ | R | 34.2949772 | -111.7874349 | 2078.13 |
| Wild Benchmark | AZ | R | 34.2973871 | -111.7856413 | 2077.85 |
| Pine Mountain - South Peak | AZ | T | 34.290341 | -111.7893836 | 2074.65 |
| Bunkerville Ridge | NV | R | 36.6593966 | -114.1288009 | 1489.34 |
| Bunkerville Ridge - Northeast Peak | NV | R | 36.6677594 | -114.1089259 | 1474.71 |
| Bunkerville Rij - W NE | NV | R | 36.6673345 | -114.1102907 | 1474.68 |
| Mica Pk | NV | R | 36.266209 | -114.1562922 | 1756.50 |
| Azure Rij | NV | R | 36.2905314 | -114.0934503 | 1478.10 |
| Azure Rij South | NV | R | 36.2764214 | -114.1021885 | 1473.28 |
| Quartzite Mtn | NV | R | 36.5017766 | -115.0875305 | 2173.35 |
| Quartzite Mtn South | NV | R | 36.498223 | -115.0881674 | 2172.71 |
| Discovery Pk | CA | T | 37.512305 | -121.699376 | 1172.41 |
| Challenger Pk | CA | R | 37.494417 | -121.678756 | 1174.55 |
| Mt Hood | CA | R | 38.4599174 | -122.5533965 | 833.84 |
| Bald Mtn | CA | R | 38.4574122 | -122.5095708 | 833.05 |
| Black Pine Mountains High Point | ID | R | 42.1384141 | -113.1254095 | 2862.32 |
| Black Pine Peak | ID | R | 42.1200078 | -113.1202851 | 2862.04 |
| Double Mountain | CA | R | 35.0333331 | -118.4867877 | 2435.63 |

| Name | State | Type | Lat | Lon | Elev |
|---|---|---|---|---|---|
| Double Mountain - East Peak | CA | R | 35.0321868 | -118.4851825 | 2434.68 |
| Peak 4191 | NV | R | 36.0454538 | -115.4312573 | 1277.54 |
| Peak 4176 | NV | R | 36.0472193 | -115.4350258 | 1272.97 |
| Blue Diamond II | NV | R | 36.0406436 | -115.4234341 | 1272.75 |
| Peak 4162 | NV | R | 36.0417331 | -115.4219823 | 1268.55 |
| Peak 4175 | NV | R | 36.0414799 | -115.4275788 | 1272.48 |
| Bear Mountain | KY | T | 37.536303 | -84.258026 | 509.11 |
| Crestone Peak | CO | R | 37.9668783 | -105.5853837 | 4358.33 |
| East Crestone | CO | R | 37.9672411 | -105.5840062 | 4358.42 |